\documentclass[a4paper,fleqn]{cas-sc}
\graphicspath{{./figures/}}

\usepackage[authoryear,longnamesfirst]{natbib}
\usepackage{algorithm}
\usepackage{algorithmic}
\usepackage{float}
\usepackage{amssymb}
\usepackage{amsmath}

\def\tsc#1{\csdef{#1}{\textsc{\lowercase{#1}}\xspace}}
\tsc{WGM}
\tsc{QE}

\begin{document}
\let\WriteBookmarks\relax
\def\floatpagepagefraction{1}
\def\textpagefraction{.001}

\shorttitle{GraphMoco}    

\shortauthors{SunRuiJin et al}  

\title {GraphMoco: a Graph Momentum Contrast Model for Large-scale Binary Function Representation Learning}  
\tnotemark[1] 


%

\author[1]{Sun runjin}


\ead{sunruijin6106@163.com}


\credit{<Credit authorship details>}

\affiliation[1]{organization={Army Engineering University of PLA},
            addressline={}, 
            city={NanJing},
            postcode={210007}, 
            state={},
            country={China}}


\author[2]{Guo Shize}

\ead{nsfgsz@126.com}


\credit{}

\affiliation[2]{organization={National Computer Network and Information Security Management Center},
	addressline={}, 
	city={BeiJing},
	postcode={100029}, 
	state={},
	country={China}}
\author[1]{Li Wei}
\ead{liwei@aeu.edu.cn}

\author[3]{Zhang XingYu}
\ead{zhangxingyu1994@126.com}

\affiliation[3]{organization={Academy of Military Sciences},
	addressline={}, 
	city={BeiJing},
	postcode={100091}, 
	state={},
	country={China}}

\author[4]{Guo Xi}
\cormark[1]
\ead{xiguo@ustb.edu.cn}

\affiliation[4]{organization={University Of Science \& Technology Beijing},
	addressline={}, 
	city={BeiJing},
	postcode={100083}, 
	state={},
	country={China}}

\author[1]{Pan ZhiSong}

\cormark[1]

\ead{hotpzs@hotmail.com}


\credit{<Credit authorship details>}





\begin{abstract}
In the field of cybersecurity, the ability to compute similarity scores at the function level for binary code is of utmost importance. Considering that a single binary file may contain an extensive amount of functions, an effective learning framework for cybersecurity applications must exhibit both high accuracy and efficiency when handling substantial volumes of data. Nonetheless, conventional methods encounter several limitations.
Firstly, accurately annotating different pairs of functions with appropriate labels poses a significant challenge, thereby making it difficult to employ supervised learning methods without risk of overtraining on erroneous labels. Secondly, while state-of-the-art (SOTA) models often rely on pre-trained encoders or fine-grained graph comparison techniques, these approaches suffer from drawbacks related to time and memory consumption. Thirdly, the momentum update algorithm utilized in graph-based contrastive learning models can result in information leakage, where the model tends to memorize the training data instead of effectively generalizing to new data. Surprisingly, none of the existing articles address this issue.
This research focuses on addressing the challenges associated with large-scale Binary Code Similarity Detection (BCSD). To overcome the aforementioned problems, we propose GraphMoco: a graph momentum contrast model that leverages multimodal structural information for efficient binary function representation learning on a large scale. Our approach employs a convolutional neural network (CNN)-based model and departs from the usage of  memory-intensive pre-trained models. We adopt an unsupervised learning strategy that effectively use the intrinsic structural information present in the binary code. Our approach eliminates the need for manual labeling of similar or dissimilar information.
Importantly, GraphMoco demonstrates exceptional performance in terms of both efficiency and accuracy when operating on extensive datasets. Our experimental results indicate that our method surpasses the current SOTA approaches such as TREX, GMN, Zeek, TikNIb, Gemini and SAFE,in terms of accuracy.
\end{abstract}


\begin{highlights}
\item At the token level, a novel cross-platform representation model named asm2vecPlus is introduced. It consists of an instruction normalization process and an instruction embedding learning network. By leveraging structural features, this model can learn instruction embeddings that embody abundant semantic information and mitigate out-of-vocabulary (OOV) problems.
\item At the block level, the multimodal CNN encoding scheme called StrandCNN is utilized to generate embeddings. This involves employing CNNs to capture the strand in the binary \citep{david_statistical_nodate}, while leveraging multiple mechanisms to optimize the model by fully utilizing the structural features of the binary. After generating the embedding, it is passed on to the high-order graph encoder for further processing.
\item 
To optimize the encoder in a graph-based contrastive learning model, a graph momentum learning model has been designed, which employs a siamese network with an embedding queue. Building a large and consistent queue allows us to train the model using less memory in large batches. 
\item The issue of information leakage is discussed. Additionally, a preshuffle mechanism has been developed to address the problem. This ensures that the model does not rely on specific patterns, but rather learns more generalized representations. 
\item We have extensively evaluated our model on multiple datasets and observed that it outperforms many SOTA models such as TREX \citep{pei_trex_nodate}, GMN \citep{Li2019GraphMN}, Zeek \citep{10.1145/3264820.3264821}, TikNIb \citep{Kim_2023}, Gemini \citep{xu_neural_2017} and SAFE \citep{massarelli_safe_2019}, across various metrics.
\item To facilitate further research, we have made our code publicly available at \url{https://github.com/iamawhalez/GraphMoco}. 
\end{highlights}


\maketitle 
{GraphMoco: a Graph Momentum Contrast Model for Large-scale Binary Function Representation Learning}

	\section{Introduction}
BCSD aims to determine the semantic consistency of a set of function pairs. As binary files are often compiled from high-level languages, files that express the same semantic meaning can vary significantly in their literal representation. BCSD finds important applications in various cybersecurity domains. For instance, in cases where library files contain vulnerable functions, BCSD can help identify them. Similarly, if patches have been applied to the code, BCSD can provide a reference to their location. When analysing malicious code, BCSD can assist in tagging it with family information.\\
The proliferation of deep neural networks has spurred an augmented adoption of these techniques within the realm of BCSD. In particular, the maturation of graph neural networks and the advancements in natural language processing have exerted a profound influence on the field. Both the academic and industrial sectors are actively involved in research efforts focused on BCSD. Nevertheless, there are several gaps that need to be addressed in the current research.\\
Firstly, One major issue is that the existing mainstream methods research rely on supervised learning models \citep{pei_trex_nodate} \citep{massarelli_safe_2019} \citep{noauthor_semantic_nodate}\citep{xu_neural_2017}, and they all suffer from the difficulty of labelling. The labeling process requires assigning labels to all functions in a set of binary functions before the training can take place. To label the distance between dissimilar function pairs, some distance functions, such as edit distance \citep{6313167}, cosine similarity \citep{massarelli_safe_2019} are commonly employed. As it is often difficult to obtain source code of binary, marking binary files often requires trained professionals and takes a lot of time. Another challenge in labeling functions arises from the fact that assigning distances of $-1$ or $0$ to dissimilar function pairs may not accurately reflect their mutual information. Researchers label the distance between the two similar functions as $\left \{ 1\right \}$ and the distance between two different functions as $\left \{-1 \right \}$ or $\left \{ 0 \right \}$. But the mutual information between any pair of functions cannot be $-1$ or $0$. For example, we might mark $\mathit{writefile()}$ and $\mathit{appendfile()}$ as $\left \{ 0 \right \}$ because they are not the same function. But in fact these two functions are very similar, the ground truth should be a score between $0$ and $1$. It is difficult for us to find a reasonable number to label each of the different function pairs. Furthermore, using integers such as $\left \{ 0 \right \}$ and $\left \{ -1 \right \}$ to label them can lead to overtraining and causing slow convergence as mentioned in \citep{ruijin2022fun2veca}.\\
Secondly, in addition to the challenges associated with labeling, the development of a practical BCSD system necessitates a delicate balance between accuracy and efficiency. Achieving higher accuracy often involves the utilization of intricate models. Nonetheless, such models tend to impose significant computational and memory demands, thereby constraining their scalability when dealing with large datasets. This is due to the fact that a single binary file can contain tens of thousands of functions \citep{280046}. The size of the functions also varies greatly, a single binary function may contain hundreds of nodes and thousands of instructions. It is challenging to balance computational costs with accuracy. When faced with large amounts of data, existing models work accurately but not efficiently. Take the existing SOTA model GMN and Order Matters \citep{yu_order_2020} for example, GMN \citep{Li2019GraphMN} makes similarity judgments about function pairs on a fine-grained comparison, which requires a large amount of computation. In the context of comparing between a new graph and all graphs within its database, the GMN model encounters the requirement of sharing node-level information concerning the new graph with all the node of all the graphs in the database. As the size of dataset increases, this process becomes computationally expensive. The Order Matters model proposed by \citep{yu_order_2020} utilizes the BERT-base model to encode node-level embeddings and sum them for the function graph. In the case of functions with a large number of nodes within a single binary, the memory consumption reaches an excessive level, posing a significant challenge. Furthermore, this problem is exacerbated by the direct correlation between the number of nodes in a function and the computational complexity and memory requirements of the model. Consequently, even though these SOTA models demonstrate remarkable accuracy, their computational and memory intensiveness restricts their practical utility.\\
Thirdly, in the realm of graph neural networks (GNNs) that incorporate contrastive learning models, the matter of information leakage has often been overlooked \citep{Qiu_2020} \citep{ren2021label}. While the original momentum contrast model \citep{DBLP:journals/corr/abs-1911-05722} successfully addressed and resolved this problem explicitly through the implementation of the shuffle mechanism, the unique data processing techniques employed in graph-based scenarios have rendered previous methods ineffective in mitigating information leakage issues. The observed behavior indicates that the GNN-based momentum contrast model exhibits a tendency to "cheat" the pretext task by quickly finding a low-loss solution. However, this expedited learning process does not translate to improved validation classification accuracy, as it deteriorates rapidly. To ensure the robustness and integrity of the model's functionality, it is crucial to find new measures aimed at mitigating potential information leakage issues.\\
We present a novel solution that not only achieves high accuracy but also demonstrates notable efficiency. Our proposed model solely relies on the utilization of Convolutional Neural Network (CNN) architecture. Inspired by recent advances in the fields of graph neural networks \citep{10.1609/aaai.v33i01.33014602}, contrastive learning \citep{DBLP:journals/corr/abs-1911-05722} and NLP, by taking full advantage of the structural features of the different layers, we propose GraphMoco: a graph momentum contrast model that uses multimodal structural information for large-scale binary function representation learning. In our proposed model, we employ a multi-level encoding strategy to represent code at the token, block, and graph levels. By adopting this approach, we aim to capture the diverse structural aspects and semantic information inherent in the code. Once the initial embeddings have been acquired, a subsequent step involves fine-tuning the encoder through the integration of contrastive learning modules.
To ascertain the efficacy of our model, extensive experimentation has been conducted. The experimental results demonstrate a significant enhancement in accuracy compared to existing methodologies, while concurrently achieving notable reductions in computational complexity and memory utilization. Additionally, the concern regarding information leakage is effectively mitigated. Consequently, we conclude that GraphMoco excels in generating high-quality binary function-level embeddings efficiently and we hope that GraphMoco will be helpful for various downstream binary analysis and graph embedding tasks.\\
In summary, the contributions we have made are as follows: \\
1. At the token level, an innovative cross-platform representation model asm2vecPlus is proposed. This model comprises two key components: an instruction normalization procedure and an instruction embedding learning network. The purpose of asm2vecPlus is to learn informative instruction embeddings that encode rich semantic information. Additionally, the model effectively addresses the issue of out-of-vocabulary (OOV) terms by leveraging the structural characteristics of the instructions.\\
2. At the block-level, deviating from the prevalent approach of utilizing pre-trained model-based architectures found in existing SOTA models, a novel and efficient multimodal CNN encoding scheme called StrandCNN is employed. The primary objective of StrandCNN is to capture the structural feature known as the strand\citep{david_statistical_nodate} within the binary code.
 \\
3. To optimize the encoder, a graph momentum learning model has been developed, making use of a siamese network architecture in conjunction with an embedding queue. In order to address the inherent problem of information leakage, a preshuffle mechanism has been incorporated into the model's design. By constructing a large and consistent queue and optimizing the model using momentum contrastive loss functions, it becomes possible to train the model efficiently with less memory. \\
4. Our model has been tested on several datasets, and the results show that it outperforms the SOTA models on several metrics. Furthermore, this model could be extended beyond BSCD application, as it can also be effectively applied in other areas of GNNs. To facilitate further research, we have made our code publicly available at \url{https://github.com/iamawhalez/GraphMoco}. 
\section{Background and Related Work}
\subsection{Definition of BCSD}
In the realm of software development, it is common practice to build upon existing code rather than creating everything from scratch. This often leads to the presence of clones in the underlying assembly code, as various source codes are frequently reused throughout the program's development. Binary Code Similarity Detection (BCSD) refers to the process of determining the level of similarity between two pieces of binary code in the absence of source code or compilation information. When presented with two pieces of binary code without access to the relevant source code or compilation details, it becomes important to ascertain their degree of similarity. This similarity, known as semantic consistency, implies that despite potential differences in the characters comprising the code, they possess identical semantics. This means that two functions, even if compiled using varying optimization techniques or originating from different architectures, will exhibit identical behavior when provided with the same input.
BCSD methods can be categorized into two distinct types: Feature-based and Embedding-based (Learning-based) approaches. Feature-based methods entail the extraction of a comprehensive set of features from binary code blocks, which are subsequently utilized to assess and quantify their similarity. Conversely, embedding-based techniques seek to learn an unsupervised, low-dimensional representation (known as an embedding) for each binary code block. These representations capture the inherent semantic information within the code. In the forthcoming sections, we will provide an elaborate description of these two methodologies.
\subsection{BCSD Based on Features}
Traditional BCSD methods heavily rely on specific features or structures. One prominent example is the EXEDIFF tool \citep{Baker99compressingdifferences}, which was introduced in 1999 to facilitate the identification of patch information in varying versions of executable files. This tool primarily leverages the structural characteristics found in disassembled code.
String-based detection methods, considered fundamental in BCSD \citep{514697} \citep{10.1145/956750.956759}, employ string operations or string edit distance measurements to detect duplicates within binary code. Another frequently employed feature is 
$n$-$grams$ \citep{khoo_rendezvous_2013}, which represent sliding windows of $N$ bytes and serve as valuable features for comparison purposes.
To mitigate the influence of different architectural variations, DiscovRe \citep{inproceedingsdis} introduces architecture-agnostic features and statistical representations for each basic block. These features are designed to be less affected by variation in architectures. However, it is important to note that while feature-based methods offer simplicity and faster processing, they may not capture all the intricate nuances present in the structures of binary code.
\subsection{BCSD Based on Embeddings}
Embedding encompasses the procedure of converting discrete variables into a continuous space with high dimensionality. Within embedding-based BCSD solutions, the assessment of functions does not rely on their inherent characteristics, but rather on the learned embeddings of these functions. Initially, the binary functions undergo transformation into multi-dimensional vector representations, commonly referred to as embeddings. Subsequently, the pairwise comparison of the two vector representations takes place through elementary geometric operations. It is observed that functions exhibiting comparable semantics possess embeddings that approximate each other in the expansive high-dimensional space. For example, a methodology proposed by \citep{massarelli_safe_2019} employs geometric operations to compare embeddings and detect similarities.\\
There exist three approaches for learning instruction-level embeddings. The first approach, known as SAFE \citep{massarelli_safe_2019}, disregards the internal structure of assembly code. Within SAFE, instructions sharing identical operands but differing opcodes are encoded as unrelated integers within their dictionaries. These dictionaries can exceed the size of 100,000. While new instructions are considered out-of-vocabulary, they are essentially amalgamations of distinct operands and operations. The second approach, introduced by \citep{li_palmtree_2021}, treats instructions as sentences and leverages BERT \citep{journalsabs181004805} to learn the embeddings. However, this model tends to consume excessive memory. The third approach, known as Asm2Vec \citep{ding_asm2vec_2019}, makes better use of the internal structure of instructions. Asm2Vec learns separate embeddings for operands and operation symbols, subsequently concatenating them together. Nonetheless, it is important to note that the Asm2Vec model is designed exclusively for x86 architecture code. \\
Drawing an analogy between assembly code instructions and words, and functions and sentences, NLP techniques can indeed be applied to solve the BCSD problem. With the remarkable success of pure attention networks in the field of NLP, such as Transformer \citep{10.5555/3295222.3295349} and BERT \citep{devlin-etal-2019-bert}, an increasing number of models are incorporating Transformer-like structures into BCSD.
In recent years, several publications, including \citep{noauthor_semantic_nodate}, \citep{li_palmtree_2021}, \citep{pei_trex_nodate}, \citep{yu_order_2020}, and \citep{Ahn2022PracticalBC}, have adopted a structure based on the pure attention pre-training model and have achieved impressive outcomes. For instance, \citet{yu_order_2020} employ a BERT-based approach, where BERT is utilized for pre-training binary code at both the one-token level and one-block levels. This demonstrates the successful application of NLP techniques, in particular BERT, to enhance BCSD performance.
Then they use convolutional neural networks (CNN) on adjacency matrices to extract the order information.  PalmTree \citep{li_palmtree_2021} adopts three pre-training tasks to capture various characteristics of assembly language and generate general-purpose instruction embeddings. These tasks are accomplished through self-supervised training on extensive unlabeled binary corpora. TREX \citep{pei_trex_nodate} introduces a novel neural architecture known as the hierarchical transformer. This architecture enables the learning of execution semantics from micro-traces during the pre-training phase, emphasizing its distinctive capabilities. By leveraging the strengths of this hierarchical transformer model, TREX aims to enhance the understanding and representation of assembly code instructions.
\subsection{BCSD Based on GNN}
With the development of Graph Neural Networks, more and more articles use GNNs to obtain the embeddings of functions. In the graph-based embedding model, the embeddings of the nodes are obtained first.
Statistical features are frequently utilized for encoding at the block level. For instance, \citet{xu_neural_2017} employ eight statistical features in their study. In a similar vein, \cite{280046} incorporates the frequency of two hundred assembly instructions. The adoption of statistical features in the existing model serves to mitigate the challenge of excessive memory demands. Nevertheless, it is worth noting that these simplistic models may give rise to potential accuracy limitations.\\
Once the node-level embedding has been acquired, the embedding of the entire function can be obtained by utilizing the Program Dependency Graph (PDG) or Control Flow Graph (CFG) information. A PDG serves as a graphical depiction of the data and control dependencies within a procedure, as described by \citep{Liu06gplag:detection}. Initially utilized in program slicing techniques as mentioned in \citep{6956589}.
 More people use CFG, such as \citep{7163056} and \citep{10.1145/2664243.2664269}. They may transform the binary code into CFGs, and use a graph analysis algorithm to find the similarity. \citet{Li2019GraphMN} propose the Graph Matching Network (GMN) in their work. GMN introduces a novel cross-graph attention-based matching mechanism to calculate a similarity score between CFGs. It achieves this by jointly reasoning on pairs of CFGs. However, for larger samples, GMN necessitates comparing each function at the node level, leading to potential challenges in terms of time complexity.\\
It is worth noting that existing models still have shortcomings in terms of accuracy or complexity when it comes to binary code similarity analysis. These limitations signify the need for further research in this domain.

\section{Problem Definition and Solution Overview}
\subsection{The Basic Definition}
In order to describe our problem clearly, we have started with a brief description of the symbols defined. Subsequently, this paper will consistently employ the following mathematical notation. The symbol $x$ represents the binary function under consideration, while $X$ denotes the dataset comprising a collection of binary functions. The function $Trans(\cdot)$ refers to the data augmentation module utilized within this framework. Correspondingly, $\vec{f_{b}}$ signifies the embedding vector associated with the block-level representation, whereas $\vec{f_{g}}$ denotes the embedding vector pertaining to the graph-level representation. In this context, the symbol $d$ serves as the distance function applied to measure the dissimilarity between two embeddings. Additionally, the function $\ell(\cdot)$represents the loss function utilized within the framework for optimization purposes.\\
\begin{table}[H]
	\caption{Mathematical Notation}
	\begin{tabular}{c|l}
		\toprule[1.5pt]
		$x$ & the\enspace binary\enspace function\enspace to\enspace be\enspace classified\\
$X$& the \enspace  binary \enspace  function \enspace dataset\\
$f(\cdot)$ & the \enspace ecoder \\
$Trans(\cdot)$& the \enspace data\enspace augmentation\enspace module \\
$\vec{f_{b}}$ & embedding \enspace  vector \enspace of \enspace  block-level \\
$\vec{f_{g}}$ & embedding \enspace  vector of \enspace graph-level \\
$d$& is \enspace the \enspace distance \enspace function \enspace of \enspace two \enspace  embeddings\\
$\ell(\cdot)$& the loss function \\
		\bottomrule[1.5pt]
	\end{tabular}
\end{table}
When two binary functions $x_{1},x_{2}$ are given, if they are compiled from the same soucecode, we say that they are similar. We aim to design a DNNs based encoder $f$. When two functions $x_{1},x_{2}$ are placed into the same encoder $f$, two different embeddings $\vec{f(x_{1}) },\vec{f(x_{2}) }$ are produced. It is desirable for a similar pair of inputs, denoted as $x_{1}$ and $x_{2}$, to exhibit proximity within a specified threshold when considering their respective embeddings, $\vec{f(x_{1}) }$ and $\vec{f(x_{2}) }$. Thus, the condition $d(\vec{f(x_{1}) },\vec{f(x_{2})}) < threshold$ hold true. Conversely, if $x_{1}$ and $x_{2}$ are dissimilar, the converse becomes evident, indicating that the condition $d(\vec{f(x_{1}) },\vec{f(x_{2})}) > threshold$ is fulfilled.
\subsection{General Framework and Solution Overview}
\begin{figure}[htbp]
	\centering 
	\includegraphics[scale=0.24]{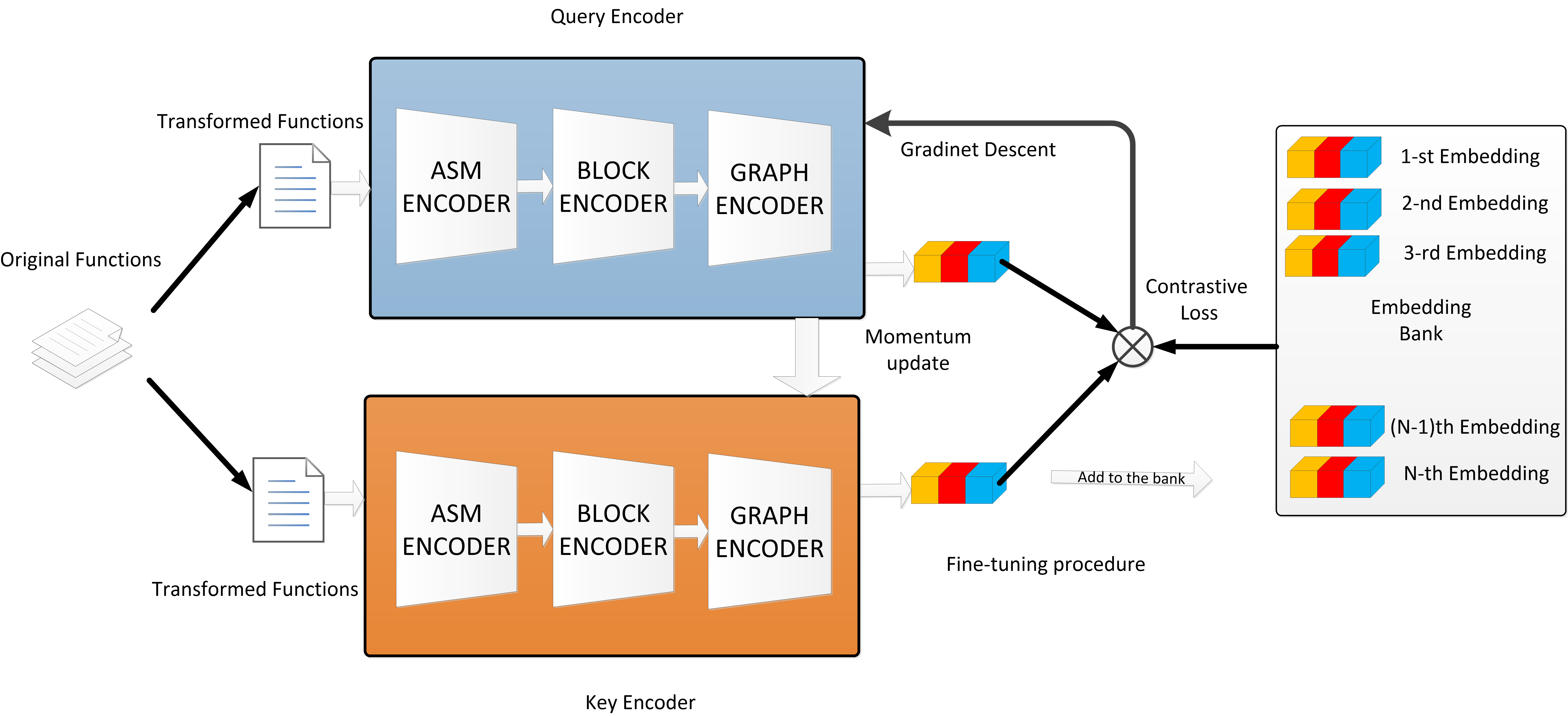} 
	\caption{The framework of the proposed unsupervised learning method, GraphMoco, comprises three main components: a data augmentation module, a siamese network encoder, and the contrastive learning module. Initially, one function is sampled. Function pairs could be produced by passing the original function to the data augmentation module. The input function pairs are processed by DNN-based encoders, resulting in a low-dimensional normalized embedding. The query encoder, referred to as $encoder_q$, is updated through back-propagation, while the key encoder, denoted as $encoder_k$, is updated using the momentum technique.
} 
	\label{fig1}
\end{figure} 
This section introduces GraphMoco, a Graph Momentum Contrast model designed to facilitate large-scale binary function representation learning by leveraging multimodal structural information. We provide an overview of the technical approach employed in this model, beginning with a description of the general algorithm. Subsequently, we provide a brief overview of the key components of the model. Notably, our model consists of a siamese graph embedding network accompanied by an embedding queue and a data augmentation module, as illustrated in Figure \ref{fig1}.
In the subsequent sections, we will delve into the technical details of the following modules:
\begin{itemize}
	\item A data set $X$ and the sample augmentation module $Trans(\cdot)$. 
 Data Set and Sample Augmentation: we define the data set as $X$, and employ a sample augmentation module denoted as 
$Trans(\cdot)$ to enhance the diversity of the samples.
 \item Instruction-level Representation Learning: we introduce Asm2VecPlus as the representation learning model operating at the instruction level. It facilitates the acquisition of meaningful representations for operands and operators, denoted as  $f_i(\cdot)$.
\item  Block-level Representation Learning: to capture representations for each block of the binary control flow graph (CFG), we utilize a block-level representation learning model known as StrandCNN. This model, denoted as 
 $f_b( \cdot )$, focuses on acquiring informative representations specific to individual blocks within the CFG.
 \item Graph-level Representation Learning: at the graph level, we utilize higher-order graph neural networks described in \cite{10.1609/aaai.v33i01.33014602}. This allows us to exploit the structural information of graph neural networks while leveraging the learned information of the graphs. We represent this graph-level representation learning model as $f_g( \cdot )$
\item Contrastive Learning Module: the contrastive learning module consists of three key components: a distance function denoted as  $d(\cdot)$, a loss function denoted as $\ell(\cdot)$, and an embedding queue referred to as $queue$. These components enable the learning of rich representations by contrasting positive and negative pairs of samples.
\end{itemize}
Due to the adoption of the contrastive learning model, there is no need for explicitly labeling dissimilar function pairs during the training process. In GraphMoco, two encoders are utilized: they are the $query$ encoder and the $key$ encoder. 
Given a function $x_{i}$ in the original dataset $X$, the augmented data could be produced by passing the original $x_{i}$ to the data augmentation module $Trans(\cdot)$. 
We could get two augmented binary functions, $x_{i}^{1}$ and $x_{i}^{2}$. The CFG of $x_{i}^{1}$ and $x_{i}^{2}$ are then produced. These augmented function-pairs are then sent to the two encoders individually, with each encoder processing one function. The query encoder $\vec{f_q()}$ is used to generate the query embeddings $\vec{f_{q}(x_{i}^{1})}$. The key encoder $\vec{f_k()}$ are used to produce the key embeddings $\vec{f_{k}(x_{i}^{2})}$. 
Each graph can be divided into blocks, and each block can be subdivided into instructions. For each instruction $\iota_{i} $, the embedding $\vec{\iota_{i}} $ are learned by the ASM encoder-Asm2VecPlus. For each block, we utilize the instruction embedding $\vec{\iota_{i}} $ along with the Block-level encoder StrandCNN to learn the block-level embedding $\vec{f_b}$, and finally the graph-level embedding $\vec{f(x_{i}^{1})}$ and $\vec{f(x_{i}^{2})}$ are obtained by using the GNNs encoder.\\
The key embeddings are added to the embedding queue. Each function instance is treated as a separate class, and a classifier is trained to distinguish between these individual instance classes. The encoded query embedding $\vec{f_{q}(x_{i}^{1})}$ should be similar to its matching key embedding$\vec{f_{k}(x_{i}^{2})}$ while being dissimilar to others in the key embeddings queue. The query encoder is updated using gradients, while the key encoder is updated through a momentum update mechanism, as described in Section 4. The contrastive learning module is then employed to optimize the encoders. Algorithm 1 provides the pseudo-code of GraphMoco.
\begin{algorithm}[ht] 
	\caption{ Framework of GraphMoco} 
	\label{alg:Framwork} 
	\begin{algorithmic}[1] 
		\REQUIRE ~~\\ 
		The set of positive samples for the i-th batch, $\left \{ x_{i}\right \} _{1}^{n}$;
		the temperature constant $\tau$; a embedding queue of $K$ keys $Q_{ueque}$;
		a siamese network of the encoder for query and key $f_q()$,$f_k()$; the data augmentation function $Trans()$, momentum update parameter $m$;
		\ENSURE ~~\\ 
		\STATE Initialize the query encoder for key encoder, the parameters of both networks are identical $f_k()=f_q()$. Initialize the embedding queque $Q_{ueque}$ with $K$ different embeddings $Q_{ueque}(k)=f(x_k)$.
 
		\FOR {each minibatch with N samples $\left \{ x_{i}\right \} _{1}^{N}$}
		\STATE Given one batch samples $\left \{ x_{i}\right \} _{1}^{N}$, two augmentated batches are generated $\left \{ x_{i}^q\right \} _{1}^{N}$, $\left \{ x_{i}^k\right \} _{1}^{N}$ with the augmentation function $Trans(\left \{ x_{i}\right \} _{1}^{N})$.
		\STATE Patch all the sub-graph $\left \{ x_{i}^q\right \} _{1}^{N}$ into one large graph $ X_{i}^q$. shuffle the sample order in the $\left \{ x_{i}^k\right \} _{1}^{N}$ mini-batch, and patch all the sub-graph of the key batch $\left \{ x_{i}^k\right \} _{1}^{N}$ into another large graph $ X_{i}^k$.
		\STATE Encode the large graph pair with the siamese network,$\vec{X^q}=\vec{f_q(X^q)}$,$\vec{X^k}=\vec{f_k(X_{i}^k)}$.
		\STATE Dispatch the $\vec{X^q}$ to get the query embedding of each samples $\vec{f_q(x_{i}^q)}$,dispatch the $\vec{X^k}$ and reshuffle sampler order to get the key embedding for each samples $\vec{f_k(x_{i}^q)}$.
		\STATE  For the current mini-batch, the queries $\vec{f_q(x_{i}^q)}$ and their corresponding keys $\vec{f_k(x_{i}^q)}$, which form the positive sample pairs. All the other samples from the queue are negative from the queries.
		\STATE Calculate the loss function $loss$ = InfoNCE($\vec{f_q(x_{i}^q)}$, $\vec{f_k(x_{i}^k)}$, $Q_{ueque}$), and update the parameters of the query encoder $f_q$.
		\STATE Update the key encoder with $f_k.params = m*f_k.params+(1-m)*f_q.params$.
        \STATE Add $N$ key embedding $\vec{f_k(x_{i}^k)}$ to the embedding queque $Q_{ueque}$.
		\ENDFOR
		\RETURN the finetuned network $f$; 
	\end{algorithmic}
\end{algorithm}

\section{Details of the Model}
In this section, we provide a comprehensive descriptions of the various components comprising the model. 
\subsection{Data Augmentation Strategies}
The role of the data augmentation module in contrastive learning is very critical. In the realm of binary code, each assembly has a specific meaning. Given a binary function code, we could produce different binaries with the same semantics. These variant binaries may be tailored for diverse architectures, subjected to obfuscation techniques, or compiled with varying optimization settings. We explore several data augmentation strategies encompassing architectural diversity, optimization variations, and bitness distinctions.\\
In our study, the selection of the CPU platform for code execution is conducted in a random manner as a general practice. Two crucial parameters considered are the CPU architectures and the CPU width. The supported CPU architectures encompass well-known mainstream types, including x86-32, x86-64, arm-32, arm-64, mips-32, and mips-64.
It is noteworthy that the assembly codes generated from a particular function written in a high-level language exhibit substantial differentiation when compiled with different options. This significant disparity arises from employing various compilation options, such as different compilers like $clang$ and $gcc$, as well as distinct optimization choices denoted by $O0,O1,O2,O3$ and $Os$. When a specific function is selected, we compile multiple instances utilizing different compilers and optimization settings.

\subsection{The Encoder}
\begin{figure}[htbp]
	\centering 
	\includegraphics[scale=0.12]{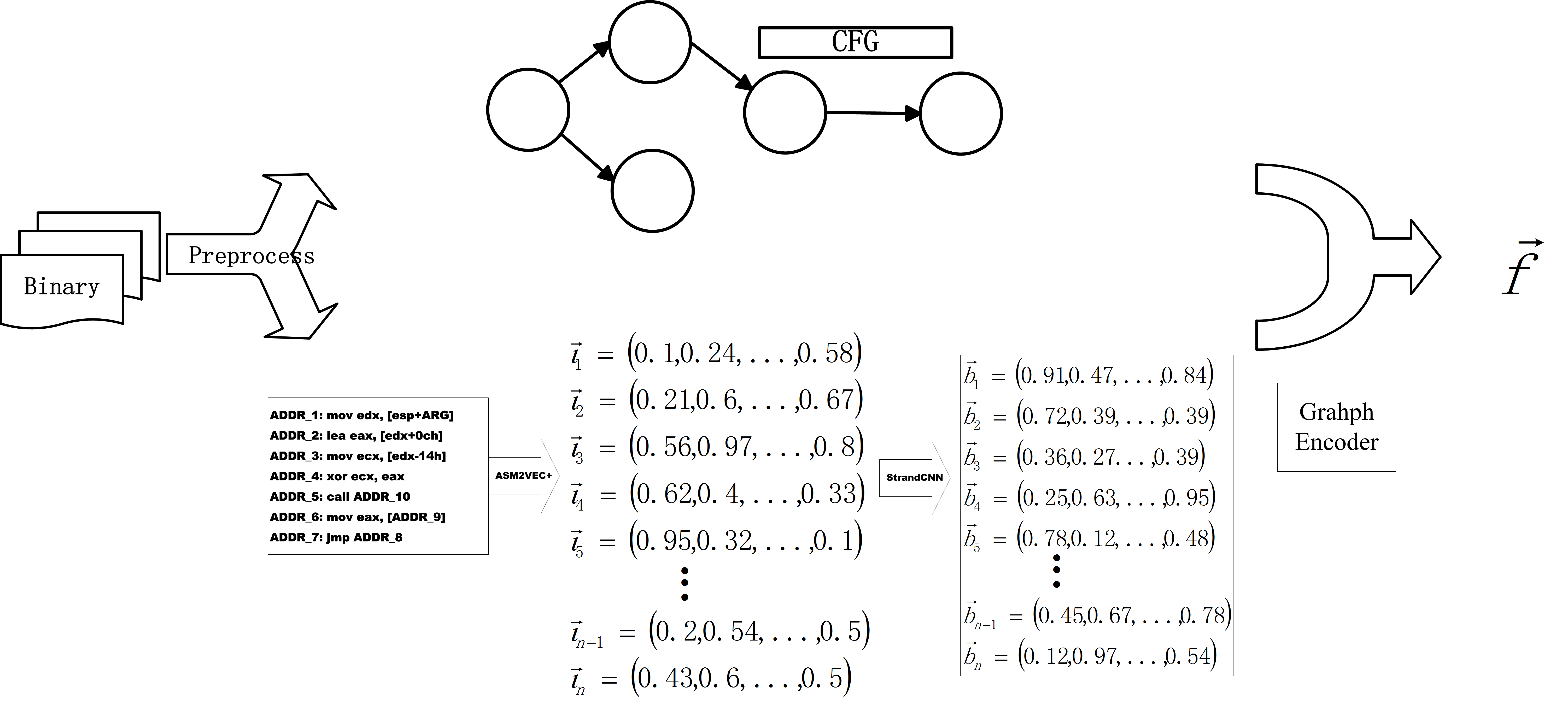} 
	\caption{The figure shows the general steps for the complete encoder. Binaries are preprocessed into ACFG. The embedding of each block are learned by the block-level encoder. Then the block-level embedding and the CFG feature are used by the graph encoder to produce the function embedding.} 
	\label{fig2}
\end{figure}

The initial step of GraphMoco involves extracting binary functions and representing them as Attributed Control Flow Graphs (ACFGs). In this representation, each vertex within the ACFG corresponds to a basic block, which is labeled with a set of instructions. Each basic block possesses a single entry point and a corresponding exit point \citep{ruijin2022instance}.
We can view the basic blocks as sentences and the instructions within each block as words. Techniques from NLP and GNNs can be applied to analyze BCSD. Specifically, Asm2VecPlus is employed to learn instruction embeddings, capturing their semantic meanings, while StrandCNN is utilized to learn block embeddings, capturing the structural information of the instructions within each block. GNNs enable the capture and assimilation of valuable information from the ACFGs. Moreover, GNNs are leveraged to learn function embeddings, enabling the extraction of higher-level representations that capture the dependencies and relationships between blocks. The encoding process of the functions is illustrated in Figure 2 (\ref{fig2}), providing an overview of the entire procedure.
\subsubsection{Asm2vecPlus:Instruction Representation Learning Module}%
\begin{figure}[htbp]
	\centering 
	\includegraphics[scale=0.24]{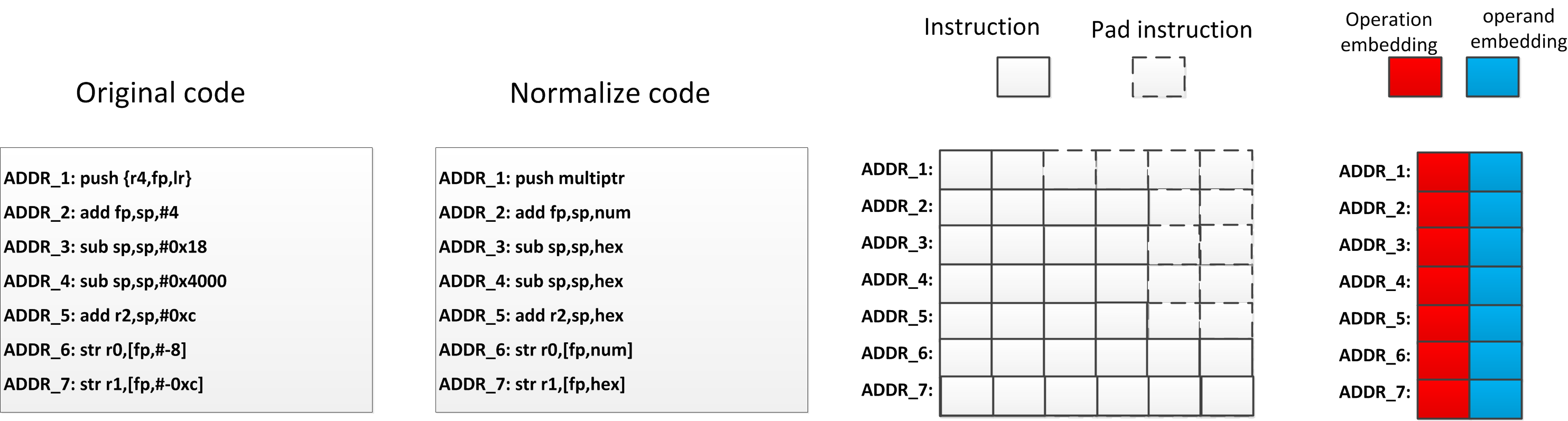} 
	\caption{The provided diagram depicts the process of the Asm2vecPlus model to generate embeddings for each instruction. The initial stage involves normalizing and padding the original codes to a fixed length. Subsequently, the operations and operands within the instructions are projected into separate embeddings. To generate the embedding for each instruction, the operand embeddings are added together and concatenated with the embedding of the operation itself. This combined embedding captures the structure feature of the instruction.} 
	\label{fig3}
\end{figure} 

To address these challenges, we developed Asm2vecPlus as an extension of the original Asm2vec architecture proposed in \citep{ding_asm2vec_2019} which is inspired by the word2vec model \citep{10.5555/2999792.2999959}. Asm2vecPlus has been enhanced to support multiple CPU architectures, providing flexibility across different systems. The overall flow of the model can be seen in Figure \ref{fig3}, illustrating the sequential steps involved in generating the instruction embeddings.\\
The initial step in our proposed solution is to associate an embedding vector with each instruction. While many researchers \citep{wang2022jtrans}\citep{pei_trex_nodate}\citep{yu_order_2020} have utilized the state-of-the-art pre-trained language model, we found that its memory consumption is substantial and it does not fully exploit the inherent lexical semantic structure of assembly instructions. Assembly instructions follow a fixed structure, typically consisting of an operation symbol and a variable number of operands. We name operands and operations in assembly code as tokens. The number of operands varies from $0$ to more than $10$. \\
In Asm2vecPlus, every instruction in the repository is mapped to a vector representation denoted as $\vec{\iota_{i}} \in \mathbb{R}^{2d} $. To address the OOV problem and ensure consistent embeddings, it is necessary to normalize the instructions before generating their vector representations. Three normalization rules are used in our article.\\
(1) To prevent the vocabulary list from becoming too large, we replace string, numbers, etc. in the assembly instruction with the special symbol \textbf{IMM}.\\
(2) The address line outside the scope of the function are replaced as \textbf{ouraddress}, and addresses within the scope of the function as \textbf{inaddress}.\\
(3) Curly brackets and all registers enclosed within them are substituted with the symbol \textbf{Pregister}. This normalization technique is particularly useful for ARM multi-register transfer instructions, as they can process any subset of the 16 visible registers with a single instruction. This replacement significantly reduces the max length of the instructions.\\
Following the normalization step, it is observed that the number of tokens encompassed within a single instruction typically spans from 1 to 5. For convenient handling, all assembly instructions are padded to a fixed length 5. Each instruction is represented by an embedding vector denoted as $\vec{\iota_{i}} \in \mathbb{R}^{2d}$. Every instruction contains a list of operands $\vec{operand_{i}^{k}} \in \mathbb{R}^{d} $ and one operation $\vec{operation_{i}} \in \mathbb{R}^{d} $. Operands and operations are treated separately in the embedding-learning process. Each operation is mapped to a numeric vector $\vec{operation_{i}} \in \mathbb{R}^{d} $. If an instruction has multiple operands ( denoted as $n$ oprands in the $i$-$th$ instruction), each operand $operand_{i}^k$, $k \in n$ can be mapped to into a numeric vector $\vec{operand_{i}^k} \in \mathbb{R}^{d} $. All the $\vec{operand_{i}^k}$ are sumed as the embeddings of operands, $\vec{operand_{i}}=\sum_{k=0}^{n} \vec{operand_{i}^k}$ and concatenated with the operation embedding $\vec{operation_{i}}$. \\
After the concatenation, the embedding of $i$-$th$ instruction is denoted as $\vec{\iota_{i}}=\vec{operation_{i}} \parallel \sum_{k=0}^{n}\vec{operand_{j}}$. By treating operands and operations separately and mapping them to numeric vectors, the Asm2vecPlus model can effectively capture both the characteristics of the operations and the individual operands, enabling it to perform well in following tasks.
\subsubsection{Block Representation Learning Module}%

\begin{figure}[htbp]
	\centering 
	\includegraphics[scale=0.15]{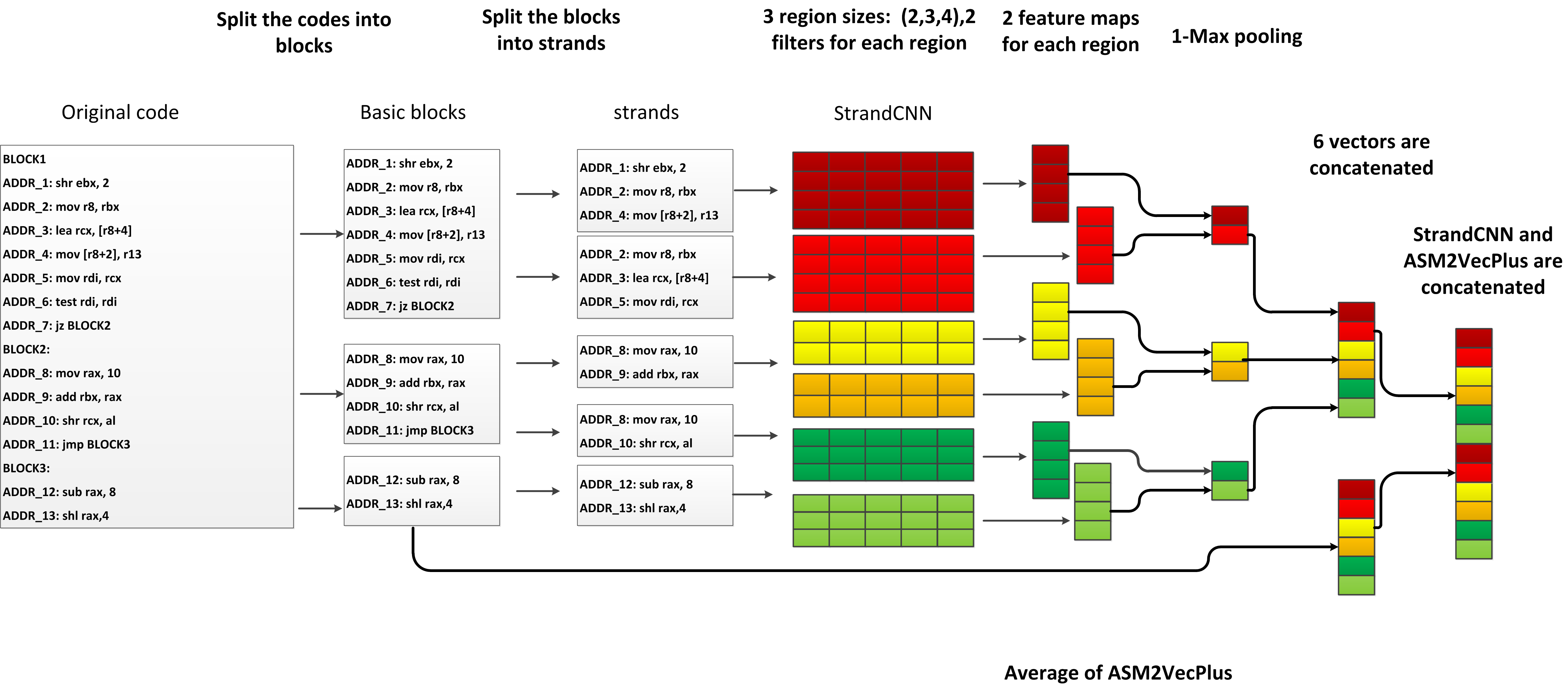} 
	\caption{Illustration of the steps involved in the StrandCNN model, a slight modification of the TEXTCNN architecture, are as follows. The model is composed of two distinct parts. In the first part, strand embeddings are acquired through convolutions of varying sizes, enabling the capture of diverse strands of the instructions. Subsequently, in the second part, the average instruction embedding is employed to encapsulate the general information pertaining to the code block. Finally, these two embeddings are concatenated to form an integrated representation.} 
	\label{fig4}
\end{figure}
With the obtained embedding representation of each instruction, we utilize it as the foundation to acquire block-level embeddings.
As depicted in Figure \ref{fig4}, the architecture of our model is named StrandCNN. It is a slight modification of the TEXTCNN approach \citep{textcnn}.
A basic block refers to a sequence of codes featuring a single entry point and a single exit \citep{ruijin2022fun2veca}. In regard to strands, they represent the set of instructions from a block that are necessary for computing the value of a specific variable \citep{david_statistical_nodate}\citep{10.1145/3264820.3264821}. By decomposing procedures into strands and utilizing them to establish similarity between procedures, we can identify a local structural feature of the binary: instructions within the same strand are usually in close proximity. We posit that learning the embedding of strands can be achieved through convolution with different filters. Drawing inspiration from \citep{textcnn}\citep{david_statistical_nodate}\citep{10.1145/3264820.3264821}, we employ CNNs to capture strands and leverage them to learn block-level embeddings.

let $\vec{\iota_{i}} \in \mathbb{R}^{2d} $ be the i-th instruction of the block. A block of length $n$ instructions is represented as
$\vec{\iota_{1}}\parallel\vec{\iota_{2}}\parallel\vec{\iota_{3}}\parallel \cdots \parallel\vec{\iota_{n}}$.
In general, let $\vec{\iota_{i:i+j}}$refer to the concatenation of instruction embeddings $\vec{\iota_{i}}\parallel\vec{\iota_{i+1}}\parallel \cdots \parallel\vec{\iota_{i+j}}$. Most of the instructions in the same strand are not far apart from each other. As shown in Figure \ref{fig4}, the length of all the strands is less than 4. We use a convolution with a filter $w \in \mathbb{R}^{h \times d} $ to produce a new feature, where $h$ is the size of the windows. We set the $h<=4$ in our article and the length is capable of learning most of the features. For example, a feature $c_{i}$ is generated from a window of instructions $\vec{\iota_{i:i+h-1}}$ by
$
c_{i}=Activate(W \cdot \vec{l}_{i:i+h-1}+b )
$.\\
Here $b \in \mathbb{{R}^{2d}} $ is a bias term and $Activate()$ is an activation functions such as the hyperbolic tangent. This filter is applied to each possible instruction window in the block $\vec{\iota_{1}}\parallel\vec{\iota_{2}}\parallel\vec{\iota_{3}}\parallel \cdots \parallel\vec{\iota_{n}}$ to produce a feature map 
$cmax[i]=max\left \{ c_{1},c_{2},c_{3},\cdots,c_{n}\right \} $.
Subsequently, a max pooling operation is applied to the feature map. The rationale behind this is to prioritize the extraction of important strands. We have outlined the procedure for extracting a single strand, and our model employs multiple filters with varying window sizes to generate multiple strands. These individual strand features are then concatenated together.
Finally, we utilize Asm2VecPlus to compute the average embedding of the block's instructions. This average embedding is then concatenated with the strand features, resulting in the final embedding representation of the block. 

\subsubsection{Function Representation Learning Module}%
Once we have obtained the block-level representations, we utilize Graph Neural Networks (GNNs) to derive the function-level embeddings. However, traditional GNNs are limited in their expressive power and cannot surpass the capabilities of the 1-WL (1-Weisfeiler-Lehman) algorithm \citep{DBLP:journals/corr/abs-1810-00826}\citep{10.1609/aaai.v33i01.33014602}. Therefore, we turn to k-dimensional GNNs (k-GNNs) as an alternative.

K-GNNs are based on the k-dimensional Weisfeiler-Lehman (WL) algorithm, which allows for the consideration of higher-order graph structures. The control flow graph (CFG) of the assembly code exhibits distinct structural features, we believe k-GNNS could learn these subgraph structures.
\subsection{Contrastive Learning Module}

In this section, we present a comprehensive exposition of the contrastive learning module, encompassing a distance function, a loss function, a siamese encoder network, and an embedding queue. The primary objective of the model is to facilitate the training of a representation encoder through the matching of an encoded query with an key embedding queue. The distance function quantify the similarity between query and the key. The contrastive loss function is used to optimize the encoder.   
\subsubsection{Siamese Network and the Embedding Queue}
When one example $x_{i}$ is sampled,  it undergoes augmentation to produce a pair of augmented functions denoted as $x_{i}^{1}$ and $x_{i}^{2}$. The augmented functions are sent to the siamese network, they are the query encoder $\vec{f()_q}$ and the key encoder $\vec{f()_k}$. The key encoder is specifically responsible for generating the key embedding. Following its creation, the key embedding is then appended to the key embedding queue. The query encoder $\vec{f()_q}$ is used to generate the query embedding $\vec{f(x_{i})_{q}}$. The query embedding $\vec{f(x_{i})_{q}}$ should be similar to its matching key embedding $\vec{f(x_{i})_{k}}$, while simultaneously demonstrating dissimilarity to all other embeddings stored within the key embedding queue. The inclusion of the embedding queue enables the comparison of the query embedding with a larger set of key embeddings, thereby facilitating training with a larger batch size without incurring excessive memory requirements. This strategy proves advantageous in achieving improved training outcomes. The calculation of the loss employs the InfoNCE (Normalized Mutual Information Neural Estimation) method, as mentioned in \citep{DBLP:journals/corr/abs-1807-03748}. The query encoder updates parameters by gradient descent and the key encoder will be updated by the momentum update mechanism as mentioned in Section \ref{momentum}.
\subsubsection{Loss Function}
In the instance discrimination framework, the assumption is made that all the samples are distinct from each other. By transforming the BSCD problem into an instance discrimination problem, a set of key embeddings $\vec{ k_{0} },\vec{k_{1}} ,\cdots,\vec{k_{n}} $ is obtained from the key embedding queue. Given an query embedding $\vec{q}$, there is only one single key that matches $\vec{q}$ ( denoted as $\vec{k_{q} } $ ). We formulate the probability that any embedding $\vec{ k_{i} }$ belongs to the $i$-$th$ function using the softmax-like criterion.
\begin{equation}\label{eq1}
P(\mathbf{k_{i}} \mid \mathbf{q})=\frac{\exp \left(\mathbf{k}_{i}^{T} \mathbf{q}\right)}{\sum_{j=1}^{n} \exp \left(\mathbf{k}_{j}^{T} \mathbf{q}\right)}
\end{equation}
where $P(\mathbf{k_{i}} \mid \mathbf{q})$ means the probability that $q$ belongs to class $i$. We also introduce a temperature parameter $\tau $ which controls the concentration level \citep{hinton2015distilling}. The probability $P(\mathbf{k_{i}} \mid \mathbf{q})$ becomes:
\begin{equation}
P(\mathbf{k_{i}} \mid \mathbf{q})=\frac{\exp \left(\mathbf{k}_{i}^{T} \mathbf{q}\right)/ \tau} {\sum_{j=1}^{n} \exp \left(\mathbf{k}_{j}^{T} \mathbf{q}\right)/\tau}
\end{equation}
We aim to maximize probability $	P(\mathbf{k_{i}} \mid \mathbf{q})$. It equivalently minimizes the negative log-likelihood $-log{	P(\mathbf{k_{i}} \mid \mathbf{q}))} $.
The loss function becomes the InfoNCE loss \citep{DBLP:journals/corr/abs-1807-03748}. The goal of fine-tuning is to minimize it.
\begin{equation}\label{eq3}
\ell_{oss}=-log{	P(\mathbf{k_{i}} \mid \mathbf{q}))}=- log\frac{\exp \left(\mathbf{k}_{i}^{T} \mathbf{q}\right)/ \tau} {\sum_{j=1}^{n} \exp \left(\mathbf{k}_{i}^{T} \mathbf{q}\right)/\tau}
\end{equation}
By minimizing the loss function $\ell$, we aim to minimize the distance between the augmented function pair. Simultaneously, we seek to maximize the cosine distance between the query function and different functions in the embedding queue. By doing so, we can achieve the objective of function classification without requiring labeled samples of different types. The loss function encourages the model to learn representations that bring similar functions close together and push dissimilar functions further apart in the embedding space, enabling effective discrimination between different functions.
\subsubsection{Momentum Update }\label{momentum}
When given the query function $q$, an ideal model would be able to accurately distinguish the corresponding key embedding $k_i$ from all the other embeddings in the queue. In other words, it would correctly identify the exact match. This scenario is equivalent to replacing the variable $n$ in Equation \ref{eq1} with the total number of embeddings in the dataset, denoted as $all$, as shown in the Equation \ref{eq4}.
\begin{equation}\label{eq4}
P(\mathbf{k_{i}} \mid \mathbf{q})=\frac{\exp \left(\mathbf{k}_{i}^{T} \mathbf{q}\right)}{\sum_{j=1}^{all} \exp \left(\mathbf{k}_{j}^{T} \mathbf{q}\right)}
\end{equation}
We can see that key features can be learned by a large dictionary that covers all the negative samples. In cases when it is infeasible to cover all examples, a higher coverage of negative samples corresponds to improved outcomes.  \\
Consequently, the crux of our approach lies in the maintenance of a large key embeddings queue. Our solution is inspired by Moco which is introduced in \citep{DBLP:journals/corr/abs-1911-05722}. 
Maintaining a embedding queue can learn good features with less memory consumption, but it also makes it intractable to update the key encoder by back-propagation. To circumvent this predicament, we employ the momentum update mechanism. \\
\begin{equation}\label{eq5}
\theta _{k} \gets m\ast \theta _{k} +(1-m) \ast \theta _{q}
\end{equation}
$\theta _{q}$ are the parameters of the query encoder, $\theta _{k}$ denote the parameters of the key encoder, and note that the two encoders are initialized with the same parameters. In each round of training, the loss is calculated by Equation \ref{eq3}. Only the parameters $\theta _{q}$ of the query encoder are updated by back-propagation. The parameters $\theta _{k} $ are updated by momentum update approach as shown in Equation \ref{eq5}. Here $m \in [0, 1)$ is a momentum coefficient, which determines the extent of momentum applied to the parameter update process.
\subsubsection{Preshuffle Mechanism}
\begin{figure}[htbp]
	\centering 
	\includegraphics[scale=0.24]{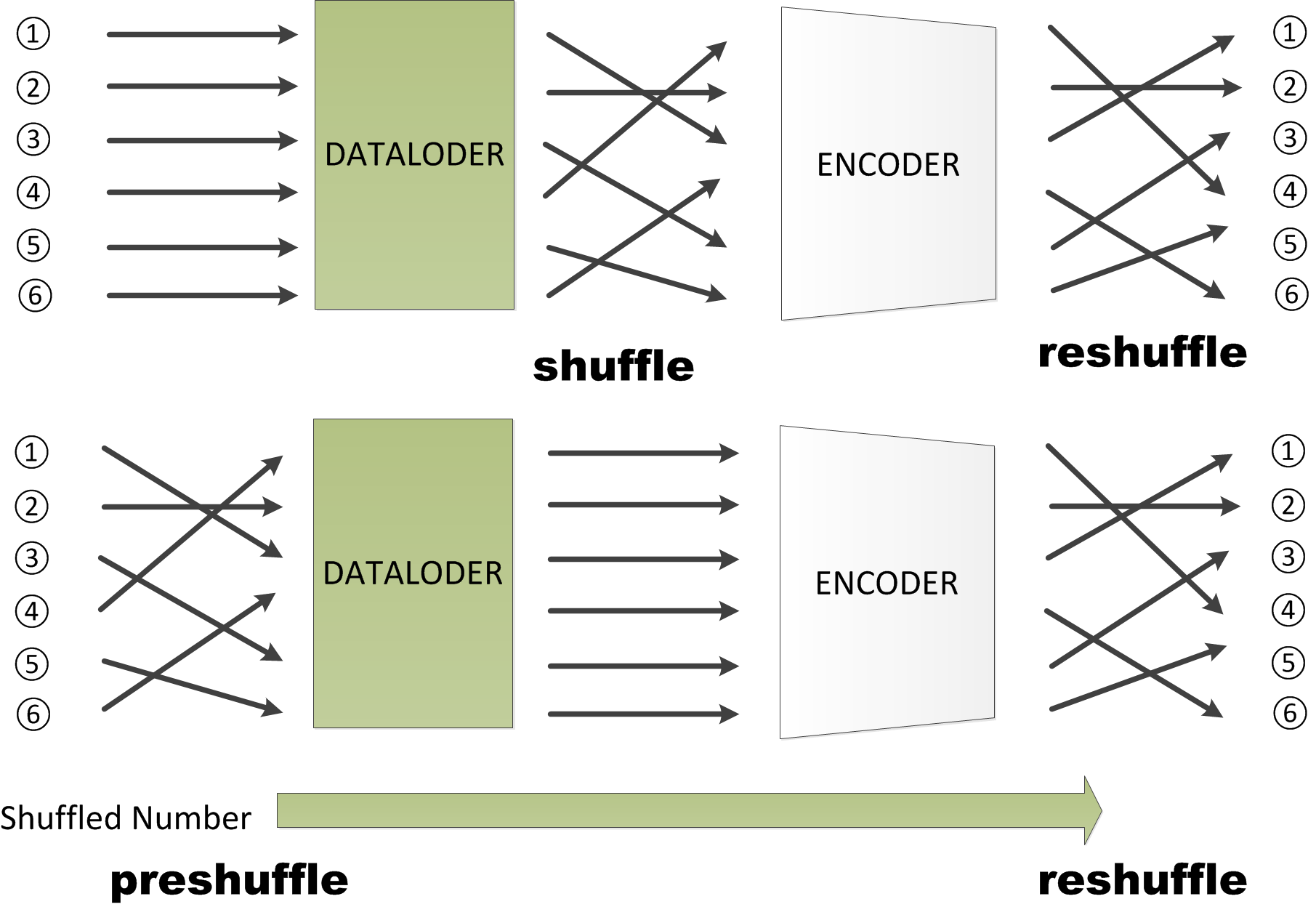} 
	\caption{ Our study examines the differentiation between two mechanisms, namely the preshuffle mechanism that we propose and the shuffle BN mechanism as introduced by \citep{DBLP:journals/corr/abs-1911-05722}. The shuffle BN mechanism involves shuffling the input data before it is transmitted to the encoder. Conversely, the preshuffle mechanism incorporates shuffling at the dataloader step. The shuffled number are added as a new parameter. } 
	\label{fig5}
\end{figure} 
Moco \citep{DBLP:journals/corr/abs-1911-05722} found that Batch Normal mechanism \citep{30451183045167} prevents the model from learning good representations. To address this issue, they introduced a shuffling procedure to disrupt Moco's ability to exploit a low-loss "cheating" solution. But the shuffling mechanism traditionally employed is not suitable for GNNs due to the requirement of reconstructing all graphs within a batch into a single graph before sending them to the encoder. When training GraphMoco, this shuffling technique proved ineffective in preventing the model from learning the cheating solution. 
To overcome these challenges, we propose a novel approach termed the preshuffle mechanism. In this method, the graphs within a batch are shuffled during the dataset packing phase, with the shuffled number being treated as a new parameter. Once the embeddings have been learned by the encoder, the batch is reshuffled again. This preshuffling technique offers an alternative and effective solution, specifically tailored for GNN-based encoders, by introducing a controlled shuffle of the graphs within the batch at the appropriate stages of the training process.
\subsubsection{Distance Function}
To measure the similarity between a pair of functions $x_{1} $ and $x_{2}$, the cosine distance function $sim()$ are employed. $\vec{f(x_{1})}$ and $\vec{f(x_{2})}$ are the embedding vectors produced by the encoder. We could use the cosine similarity as the distance metric $d(x_{1},x_{2})$. The cosine similarity formula is shown below:
\begin{equation}
d(x_{1},x_{2})=sim(x_{1},x_{2})=\frac{\vec{f(x_{1})} ^{\top} \vec{f(x_{2})}}{\left \| \vec{f(x_{1})}\right \| \left \| \vec{f(x_{2})}\right \| } =\frac{\sum_{i=1}^{d} \left (\vec{f(x_{1}) }\left [ i \right ] \cdot \vec{f(x_{2}) }\left [ i \right ] \right ) }{\sqrt{\sum_{i=1}^{d} \vec{f(x_{1})}\left [ i \right ]^{2} } \cdot \sqrt{\sum_{i=1}^{d} \vec{f(x_{2})}\left [ i \right ]^{2} } }
\end{equation}
$\vec{f(x)}\left [ i \right ]$ indicates the $i$-$th$ component of the embedding vector of the function $x$. The cosine distance between two function embeddings is $[-1,1]$. Where 1 means exactly the same, -1 means completely different. \\

\section{Experiments and Evaluation}
The evaluation of GraphMoco encompasses various tasks to assess its performance across different scenarios:\\
Task 1 - Cross Platform Tests On Dataset1: This test aims to evaluate the accuracy of GraphMoco in BCSD tasks with different methods. The evaluation is conducted on a publicly available dataset, focusing on assessing how well GraphMoco performs compared to other techniques.\\
Task 2 - Function search Tests On Dataset2: In the second task, we increased the difficulty by more than 10 times. Given a specific binary function, the objective is to identify similar functions within a large dataset generated by multiple compilers. GraphMoco's performance in this function search task is assessed.\\
Task 3 - Vulnerability Search on Dataset3: In this task, GraphMoco is evaluated on a real-world scenario. We search for vulnerable functions from the firmware that contain vulnerabilities .\\
Task 4 - Ablation Analysis:  This section focuses on quantifying the contributions of different components of GraphMoco to the final results. By conducting ablation experiments, we could understand the impact and significance of various parts of the GraphMoco model.\\
These extensive evaluations across different tasks provide insights into the performance and capabilities of GraphMoco in diverse application scenarios related to code similarity detection, function search, vulnerability identification, and understanding the importance of different model components.

\subsection{Setups 
	\label{sec:setups}}
\textbf{Setup}. The experimental environment was established on a Linux server running Ubuntu 20.04. The server configuration consisted of two Intel Xeon 4210r CPUs clocked at 2.4 GHz, with each CPU containing 10 physical cores and 20 virtual cores. Moreover, the server was equipped with 128GB of memory. To leverage GPU acceleration, a single Nvidia RTX 3090 GPU with 24GB of memory was employed. The software environment encompassed Python 3.8 and PyTorch 1.11.0, exploiting the capabilities of CUDA 11.7 for accelerated computations.\\
\textbf{Hyperparameters}. We implemented the GraphMoco model with the following settings. The model was fine-tuned for a total of 100 epochs, using a temperature value of 0.07 and an output embedding dimension of 256. The learning rate was set to 0.01, and we employed the Adam optimizer for training. During training, a weight decay of $1e$-$4$ was applied, and a batch size of 128 was used. The embedding queue size was set to 5120, and a momentum value of 0.999 was assigned. These settings were chosen to optimize the performance and stability of the model during the training process.\\
\textbf{Metrics}. In our practical implementation, we leverage function pairs that are compiled from the same source code and as similar ones. This approach saves a lot of manual labelling effort, and it is also a common practice in academia \citep{ding_asm2vec_2019} \citep{pei_trex_nodate}. To quantify the similarity between pairs, we utilize the cosine similarity function. It measures the distance between the embedding vectors of two functions, allowing us to assess their degree of similarity effectively. 

\subsection{Task1: Cross Platform Tests On Dataset1}
Dataset-1 consists of seven  widely-used open-source projects: ClamAV, Curl, Nmap, OpenSSL, Unrar, Z3, and Zlib. Once compiled, these projects generate a total of 24 distinct libraries. Each library is compiled using two compiler families, GCC and Clang, each with four different versions, covering major releases from 2015 to 2021. More details on the opensource projects and compiler versions are provided in \citep{280046}. Additionally, each library was compiled for three different architectures, x86-64, ARM, and MIPS, encompassing both 32-bit and 64-bit versions. This approach resulted in a total of six architecture combinations, and 5 optimization levels O0, O1, O2, O3, and Os. Based on this setup, we identified and evaluated seven distinct tasks, each associated with a specific combination of project, compiler version, architecture, and optimization level. These tasks allowed us to comprehensively assess the performance and characteristics of the compiled libraries under different configurations and settings: 
(1) arch: the function pairs have different architectures. 
(2) opt: the function pairs have different optimizations, but the same compiler, compiler version, and architecture. 
(3) comp: the function pairs have different compilers. 
(4) XC: the function pairs have different compilers, compiler versions, and optimizations, but the same architecture and bitness. 
(5) XC+XB: the function pairs have different compiler, compiler versions, optimizations, and bitness, but the same architecture. 
(6) XA: the function pairs have different architectures and bitness, but the same compiler, compiler version, and optimizations. 
(7) XM: the function pairs are from arbitrary architectures, bitnesses, compilers, compiler versions, and optimization levels. \\
In our methodology, the employment of cosine distance serves as a robust metric to ascertain their equivalency between two functions. If the distance between a pair of functions falls below a certain threshold, we consider the functions to be different. However, we don't rely solely on a specific threshold to evaluate the robustness of the detection model. Instead, we employ alternative metrics, such as the AUC-ROC (Area Under the Receiver Operating Characteristic Curve) mechanism, MRR10 and RECALL@1.
The AUC-ROC metric is commonly used to assess the performance of binary classification models. It measures the model's ability to distinguish between positive and negative samples by calculating the area under the ROC curve. A higher AUC-ROC value indicates better performance in terms of correctly classifying similar and dissimilar function pairs. MRR evaluates how well the system ranks the correct or relevant item at the top of the list. It considers the reciprocal rank of the first relevant item found for each query, and then calculates the average reciprocal rank across all queries. Recall@1 calculates the percentage of queries where the first result in the ranked list is a relevant item. It indicates how well the system is able to retrieve the most relevant item as the top recommendation.
For the most complex XM task, we employ three different metrics to evaluate the results: AUC, MRR10, and RECALL@1. For other tasks we use only AUC.\\
\begin{table}[H]	 			
	\caption{Results on Function Pairs across Architectures, Optimizations, and Obfuscations.}
	\centering
	
	\resizebox{\linewidth}{!}{
		\begin{tabular}{c ccc ccc ccc cc}
			\toprule[1.5pt]
			\multirow{2}{*}{index} & \multirow{2}{*}{model name}& \multirow{2}{*}{Description} & \multirow{2}{*}{XA}& \multirow{2}{*}{XC}&\multirow{2}{*}{XC+XB}& \multirow{2}{*}{XM}& \multirow{2}{*}{arch}	&\multirow{2}{*}{comp}&\multirow{2}{*}{opt}&\multicolumn{2}{c} {XM}\\ 
			&&&&&&&& && {MRR10}&{Recall@1}\\
			\midrule
		1&	GraphMoco&with StrandCNN	& \textbf{0.88}&\textbf{0.88}&\textbf{0.89}&\textbf{0.89}& \textbf{0.99}&0.81 &\textbf{0.99}&\textbf{0.60}&\textbf{0.55}\\
			\midrule
		2&	GMN&cfg		&0.82&0.82&0.83&0.82&	0.93&0.79	&0.84	&0.37&0.29\\
		3&	GMN&OPC-2000 			&0.86&0.86&0.87&0.87&	0.97&0.80	&0.88	&0.52&0.44\\
			
			\midrule
		4&	GMN&  opc200 	&0.86&0.86&0.87&0.87&	0.97&\textbf{0.82}	&0.89	&0.53& 0.45 \\
		5&	GMN& 	cfg    		&0.86&0.85&0.86&0.86& 	\textbf{0.99}&0.77	& 0.89	& 0.43& 0.33 \\		
			
			\midrule
		6&	Gemini& 	GeminiFeatures	&0.80&0.81&0.82&0.81&	0.96&0.72	&0.84	&0.36& 0.28\\		
		7&	Gemini& NoFeatures 			&0.69&0.69&0.70&0.70&	0.76&0.66	&0.69	&0.12& 0.07\\		
		8&	Gemini& OPC-200 			&0.78&0.81&0.82&0.81&	0.93&0.74	&0.85	&0.369& 0.26 \\
			\midrule
		9&	GNN	& 	ArithMean 			&0.74&0.79&0.79&0.77&	0.88&0.73	&0.83	&0.25& 0.16 \\		
		10&	GNN&	AttentionMean		&0.76&0.79&0.79&0.77&	0.88&0.72	&0.80	&0.29& 0.20 \\
			\midrule
		11&	SAFE& ASM-list\_150 	&0.82&0.82&0.83&0.83&	0.94&0.76	&0.83	&0.22& 0.09 \\
		12&	SAFE& ASM-list\_250   	&0.79&0.79&0.80&0.80&	0.94&0.70	&0.81	&0.28& 0.17 \\
		13&	SAFE& ASM-list\_Trainable\_e10 			&0.80&0.80&0.80&0.81&	0.95&0.71	&0.82	&0.29& 0.15 \\
			\midrule
		14&	Trex & 	512tokens 			&0.73&0.80&0.79&0.77&	0.87&0.80	&0.89	& 0.31& 0.25\\
		15&	PVDM & 	 10 CFG random walks 		&0.62&0.81&0.74&0.69&	0.74&0.92	&0.96	& 0.14& 0.09 \\
			\bottomrule[1.5pt]
			
	\end{tabular}}
	\label{tab:tabfirst}
\end{table}
The results of our models are reported in Table \ref{tab:tabfirst}.
In our analysis, we have picked six distinct models for comparative evaluation alongside our own model. These models encompass the TREX\citep{pei_trex_nodate}, GMN\citep{Li2019GraphMN}, PV-DM\citep{le_distributed_nodate}, Gemini\citep{xu_neural_2017}, SAFE\citep{massarelli_safe_2019}, and Zeek\citep{10.1145/3264820.3264821} architectures. Furthermore, careful consideration has been given to employing diverse encoders specifically tailored to each individual model.
The indexing scheme is as follows: Index 1 signifies the outcome attained through the application of our GraphMoco model with StrandCNN. Indices 2 and 3 correspond to the GMN model, incorporating a direct comparison approach. CFG features and 200 opc features are used. Indices 4 and 5 encapsulate the GMN model leveraging an indirect comparison methodology. CFG features and 200 opc features are used. Direct comparison entails a straightforward assessment of similarity between two functions, where the degree of resemblance is directly provided. Conversely, in indirect comparison, the representations of two functions are acquired through a process of learning embeddings, which in turn enables the utilization of cosine distance as a metric for measuring their similarity. Indices 6, 7, and 8 pertain to the Gemini model, highlighting the exclusive utilization of CFG Features, Gemini Features, and 200 OPC features respectively. Indices 9 and 10 showcase the application of the Structure2vec model, featuring unsupervised ArithMean and AttentionMean embedding. Indices 11, 12, and 13 represent the SAFE model. They employ the maximum instruction length of 150, the maximum instruction length 250, and trainable embedding features respectively. Index 14 showcases the employment of the Trex model with 512 tokens, while index 15 showcases the utilization of the PVDM model with 10 CFG random walks.\\
 Notably, GraphMoco exhibits superior performance compared to all existing models across most evaluation tests, with the exception of the "comp" task. Particularly noteworthy is the exceptional accuracy achieved by GraphMoco, nearly reaching 100\%, for the simplest task "arch". There is little difference between GraphMoco and the existing SOTA model in terms of the AUC metrics in most task. However, our models demonstrate a significant competitive edge over other SOTA models when considering the MRR10 and recall@1 metrics. Specifically, GraphMoco attains scores of 0.57 and 0.51. Among all the models considered, the GMN model exhibits the highest degree of similarity to our own model. Notably, the GMN \citep{Li2019GraphMN} model yields scores of 0.52 and 0.45, respectively. These outcomes suggest that GraphMoco excels, particularly in intricate tasks.
\subsection{Task2: Function Search Tests}
In Task 1, our investigation of the model's capabilities encompassed a comprehensive evaluation. However, it is crucial to acknowledge that the previous task was conducted utilizing relatively modest poolsizes, approximately 100 functions. This approach facilitated the identification of corresponding functions within this limited scope. In light of the recognition that real-world scenarios frequently involve substantially larger poolsizes, we deliberately shifted the number in Task 2. The primary objective of Task 2 was to enhance the difficulty level, thereby more accurately reflecting the challenges encountered in practical applications. By doing so, our investigation elevate the model's robustness and efficacy when confronted with the complexities inherent in analyzing considerably larger pools of functions.
The objective of Task 2 was to identify ten similar functions that exhibit semantic identity with the target function from a pool of 10,000 functions. It is worth highlighting that the complexity of Task 2 has substantially heightened compared to Task 1, thereby posing a more challenging scenario.
The dataset is adapted from a big data competition run by 360 and National Engineering Research Center for Big Data Collaborative Security Technology
\footnote{https://www.datafountain.cn/competitions/593}. This dataset comprises a total of 70,000 functions, each uniquely identified by a function ID. These functions were compiled using three distinct compilers, namely gcc, llvm, and clang, across three different CPU architectures and two bit types. Additionally, five distinct levels of optimization were applied during the compilation process.
The dataset consists of over 2.4 million compiled samples, with an average of more than thirty samples per function. To facilitate model training and evaluation, we partitioned the dataset into three subsets: a training set, a validation set, and a test set. The training set comprises 50,000 functions, encompassing approximately 1,600,000 samples. The validation set contains 3,000 functions, consisting of 100,000 samples. Lastly, the test set consists of 17,000 functions, with a total of 500,000 samples.
To assess the performance of our models, we employed the Mean Average Precision (MAP) metric. Average Precision (AP) represents the average accuracy of an individual prediction result, as described by Equation 7.
\begin{equation}
AP=\frac{\sum_{k=1}^{n}(P(k) \times \operatorname{rel}(k))}{N_{rel}}
\end{equation}
where $n$ is the total number of prediction functions, and $n$ is set to 10 in this task; $k$ is the ranking position of the prediction functions; $P(k)$ is the percentage of the first $k $ prediction functions that are truly relevant; $rel(k)$ indicates whether the $k_th$ prediction function is relevant, and relevant is $1$, and vice versa is $0$; $Nrel$ indicates the total number of truly relevant functions in the test dataset. MAP is the mean value of the average accuracy of the multiple prediction results, where $q$ is the serial number of the prediction, $Q$ is the total number of predictions, and AP$(q)$ is the average accuracy of the $q_th$ prediction result.
\begin{equation}
MAP=\frac{\sum_{q=1}^{Q} A P(q)}{Q}
\end{equation}
\begin{table}[H]
	\caption{Results of Task 2 On Test Dataset}
	\centering
	
	
	\begin{tabular}{ccc}
		\toprule[1.5pt]
		model name & Description &MAP\\
		\midrule
		GraphMoco& StrandCNN&0.91\\
		\midrule
		Gemini& Gemini features&0.18\\
		Gemini& CFG +bow opc 200 &0.23\\
		
		Gemini &StrandCNN &0.35\\
            \midrule
		GATv2 & StrandCNN&0.45\\
		\midrule
		TREX& 512tokens&0.76\\
		Func2vec& 512tokens&0.86\\
		\midrule
		SAFE& ASM-150& 0.55\\
		\midrule
		TikNIb\citep{Kim_2023}&Interpretable Feature&0.18\\
		\bottomrule[1.5pt]
		\label{tab360}
	\end{tabular}
	
\end{table}
In our analysis, we have picked six distinct models for comparative evaluation alongside our own model. These models encompass the TikNIb\citep{9813408}, SAFE, Func2vec\citep{ruijin2022fun2veca}, TREX, Gemini, and GATv2\citep{Brody2021HowAA} architectures. 
The increased difficulty of Test 2 posed a significant challenge for the existing SOTA models, a significant number of models struggled to perform satisfactorily. Only the bert-based model TREX \citep{pei_trex_nodate} and Func2Vec \citep{ruijin2022fun2veca} managed to achieve a score above 0.6. However, GraphMoco stood out by achieving an impressive score of 0.91, demonstrating its superior capability to handle the enhanced difficulty level.
\subsection{Task3: Vulnerability Search}

\begin{table}[H]
	\caption{Vulnerabilities We Have Confirmed(\checkmark ) In Firmware Images}
	\centering
	
	\resizebox{\linewidth}{!}{ 
		\begin{tabular}{cccccc|}
			
			\toprule[1.5pt]
			CVE 			&Description&function name 				&TP-Link Deco-M4 &NETGEAR R7000\\
			\midrule
			CVE-2019-1563	&Decrypt encrypted message&MDC2\_Update				&\checkmark &\checkmark\\
			CVE-2019-1563	&Decrypt encrypted message& PKCS7\_dataDecode			&\checkmark&\checkmark\\
			
			CVE-2016-6303 	&Integer overflow& MDC2\_Update				&\checkmark& \\
			CVE-2016-2182 	&Allows denial-of-service& BN\_bn2dec				&\checkmark& \checkmark\\
			\midrule
			CVE-2016-2176 	&Buffer over-read& X509\_NAME\_oneline		& &\checkmark\\
			
			CVE-2016-2106	&Integer overflow&EVP\_EncryptUpdate		& &\checkmark\\
			CVE-2016-2105	&Integer overflow&EVP\_EncodeUpdate			& &\checkmark\\
			CVE-2016-0798	&Allows denial-of-service&SRP\_VBASE\_get\_by\_user	& &\checkmark\\
			CVE-2016-0797	&NULL pointer dereference&BN\_dec2bn				& &\checkmark\\
			CVE-2016-0797	&NULL pointer dereference&BN\_hex2bn					& &\checkmark\\
			\bottomrule[1.5pt]
			\label{tab2}
	\end{tabular}}
\end{table}
In this task, we aim to evaluate our model's performance in a real-world scenario involving the identification of n-day vulnerabilities within firmware images. Firmware images commonly contain numerous third-party libraries, but when these libraries are patched, the corresponding firmware updates are often delayed or not released at all.
For this evaluation, we utilize a dataset provided by \citep{280046}. The dataset comprises ten vulnerable functions carefully selected from OpenSSL1.0.2d, which collectively cover eight CVEs (Common Vulnerabilities and Exposures). The target functions we focus on belong to the libcrypto libraries embedded within two specific firmware images: Netgear R7000 (ARM 32-bit) and TP-Link Deco M4 (MIPS 32-bit). These firmware images serve as representative examples for our analysis.
The Netgear R7000 firmware exhibits four distinct vulnerabilities, while the TP-Link Deco M4 firmware contains nine vulnerabilities. In order to provide a comprehensive analysis, we present a detailed account of these vulnerabilities. Our analysis involves the extraction of 1,708 functions from the TP-Link Deco M4 firmware and 1,776 functions from the Netgear R7000 firmware. Subsequently, we compare these extracted functions with the aforementioned vulnerabilities. The Mean Reciprocal Rank at K (MRR10) serves as the metric for comparison.\\
In our analysis, we have picked six distinct models for comparative evaluation alongside our own model. These models encompass the TREX, GMN, PVDM, Gemini, SAFE, and Zeek architectures. Furthermore, for GMN, Gemini, SAFE, we employ diverse encoders specifically tailored to these model.
The obtained results indicate a significant superiority of our model over the existing SOTA model in the majority of cases (6 out of 8), resulting in an average improvement of approximately 9\%. However, our model demonstrates a marginally lower performance compared to the SOTA model in a few instances (2 out of 8), leading to an average reduction of approximately 2\%. 

\begin{table}[H]
	\caption{Results MRR10 of Vulnerability test.}
	\centering
	
	\resizebox{\linewidth}{!}{
		\begin{tabular}{ccc ccc ccc c}
			\toprule[1.5pt]
			\multirow{2}{*}{model name}& \multirow{2}{*}{Description} & \multicolumn{4}{c} {Netgear R7000}& \multicolumn{4}{c} {TP-Link Deco-M4}\\ 
			&&x86&x64&ARM32&MIPS32&x86&x64&ARM32&MIPS32\\
			\midrule
			GraphMoco&with TEXTCNN	& \textbf{0.88}&\textbf{0.63}&\textbf{1}&\textbf{0.88}	& 0.65&\textbf{0.92} &0.67&\textbf{0.85}\\
			\midrule
			GMN& cfg+bow opc200		&\textbf{0.88}&0.54&\textbf{1}&0.79		&\textbf{0.67}&0.73	&\textbf{0.70}	&0.78\\
			GMN& 	cfg 			&0.65&0.43&0.69&0.54				&0.80&0.88	&0.88	&0.98\\
			
			
			\midrule
			GNN-s2v\_GeminiNN&GeminiFeatures&0.33&0.042&0.38	&0.25&0.11&0.26&0.28&0.11\\
			GNN-s2v\_GeminiNN&NoFeatures&0.0&0.0&0.028&0.0		&0.0&0.0&0.11&0.0\\
			GNN-s2v\_GeminiNN&OPC-200&0.33&0.31&0.67&0.34		&0.39&0.28&0.36&0.59\\
			
			\midrule
			GNN-s2v&ArithMean\_e5&0.1&0.08&0.5&0.0		&0.05&0.06&0.028&0.18\\
			GNN-s2v&AttentionMean\_e5&0.05&0.03&0.04&0.0	&0.0&0.03&0.04&0.27\\
			
			\midrule
			SAFE&ASM-list\_250\_e5&0.0&0.0&0.0&0.0	 &0.04&0.02&0.07&0.03\\
			SAFE&ASM-list\_Rand\_Trainable\_e10&0.0&0.0&0.0&0.03 &0.06&0.16&0.11&0.07\\
			SAFE&ASM-list\_Trainable\_e10&0.08&0.29&0.29&0.16 &0.04&0.16&0.24&0.09\\
			\midrule
			Trex & 	512tokens&0.40&0.48&0.44&0.5	&0.29&0.42&0.22&0.61\\
			asm2vec & 	e10 &0.25&0.13&0.86&0.5	&0.11&0.11&0.02&0.67 \\
			
			\bottomrule[1.5pt]
			\label{tab5}
	\end{tabular}}
\end{table}

\subsection{Ablation}
In this section, we aim to quantify the contributions of different components of GraphMoco towards the final result. To achieve this, we evaluate three tasks: the preshuffle mechanism, the contrastive loss, and the effect of the StrandCNN encoder.
\subsubsection{The Preshuffle Mechinism}
\begin{figure}[htbp]
	\centering 
	\includegraphics[scale=0.5]{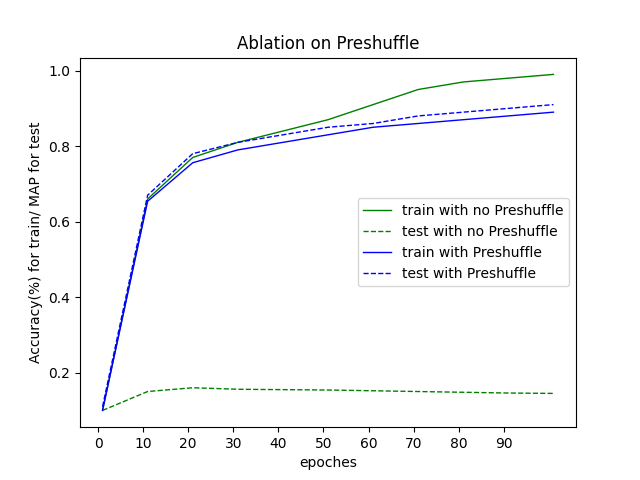} 
	\caption{The figure presented in this study illustrates the training and test curves of GraphMoCo, both with and without the preshuffle BN mechanism. In this representation, the blue color is used to depict the curves obtained when employing the preshuffle mechanism, while the green color corresponds to the curves obtained without the preshuffle mechanism. Notably, the dotted line in the figure represents the performance specifically on the test set.
} 
	\label{figpr}
\end{figure}

Firstly, we examine the preshuffle mechanism. This mechanism plays a crucial role in randomizing the input graph structures, thereby enhancing the model's ability to capture diverse patterns and generalize well.
To illustrate the impact of the preshuffle mechanism, we conducted an experiment where we trained Task 2 by excluding the preshuffle mechanism. Figure \ref{figpr} presents the training and test curves of GraphMoCo with and without the preshuffle mechanism. In the diagram, the dashed line represents the performance without the preshuffle mechanism, while the blue color indicates the performance on the test set.

From the figure, it is evident that removing the preshuffle mechanism leads to noticeable overfitting to the pretext task. The training accuracy of the pretext task rapidly reaches 99\%, indicating that the model quickly learns to excel in the training data. However, this high accuracy does not translate to good generalization on the test set, as the test accuracy experiences a significant drop. In contrast, when training with the preshuffle mechanism, the improvement in accuracy is more gradual. Despite this slower progress during training, the model still achieves competitive results on the test set. This demonstrates the significance of the preshuffle mechanism in preventing the model from learning deceptive patterns or "cheating" results.

Overall, these findings emphasize the importance of the preshuffle mechanism in GraphMoCo. It not only helps mitigate overfitting to the pretext task but also ensures that the model can generalize well to unseen data, thus enhancing its overall performance and reliability.
These findings indicate that without the preshuffle mechanism, the model becomes overly reliant on some statistical cues within each batch. In contrast, the presence of the preshuffle mechanism prevents such information leakage by disrupting the statistical patterns, thereby ensuring improved generalization and preserving the model's ability to learn meaningful representations.
\subsubsection{The Effect of the Contrastive Learning }	
In order to evaluate the performance of GraphMoco, we conducted a comparative analysis with two traditional models: the supervised learning model and the triplet loss model \citep{Schroff_2015}. The evaluation involved encoding functions using the same encoder employed in GraphMoco and fine-tuning the models using different methods.

For the supervised learning model, we maintained a balanced ratio between similar and dissimilar function pairs in the fine-tuning set, approximately 1:1. This rate follows the setting of \citep{pei_trex_nodate}. The testing was performed on Dataset2, following the same experimental settings as Task 2. The triplet loss model showcased performance with a MAP score of 0.8.

The results indicate that our proposed model achieved a significantly higher accuracy compared to the traditional models. Specifically, the accuracy of GraphMoco surpassed the traditional models by 0.11 points. The GraphMoco showcased performance with a MAP score of 0.91. These findings provide empirical evidence of the efficacy of GraphMoco in improving the accuracy of function encoding and demonstrate its superiority over conventional approaches.

\begin{figure}[htbp]
	\centering 
	\includegraphics[scale=0.5]{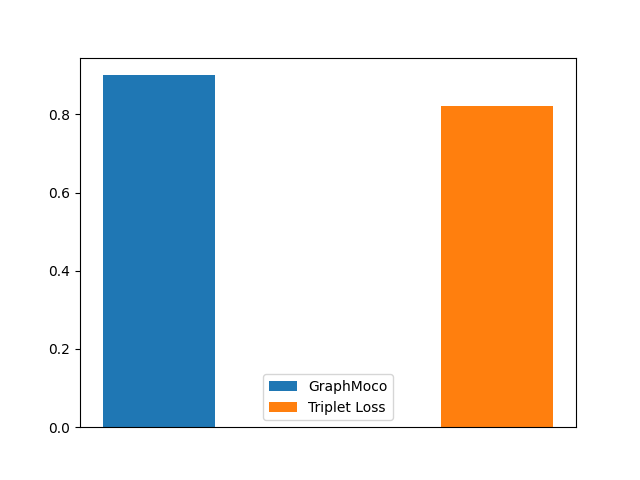} 
	\caption{The impact of contrastive learning was assessed by comparing our model to the triplet loss approach. Our findings revealed that our model exhibited a notable improvement in MAP, outperforming the triplet loss approach by approximately 10 percent. } 
	\label{fig6}
\end{figure} 

\subsubsection{The effect of the StrandCNN encoder }	
To further investigate the impact of the StrandCNN encoder, a comparative analysis was conducted to examine its performance in relation to other encoders. The findings revealed that the utilization of the StrandCNN encoder exhibited superior effectiveness. Notably, the observed improvement over FastText--avearage of 1-gram,2-gram and 3-gram  embedding \citep{DBLP:journals/corr/JoulinGBM16} is obvious.  Furthermore, a comparative evaluation was performed between the results obtained using the StrandCNN encoder and those achieved by solely employing ASM2VecPlus.
\begin{table}[H]
	\caption{Results on function pairs across architectures, optimizations, and obfuscations.}
	\centering
	
	\resizebox{\linewidth}{!}{
		\begin{tabular}{ccc ccc ccc cc}
			\toprule[1.5pt]
			\multirow{2}{*}{model name}& \multirow{2}{*}{Description} & \multirow{2}{*}{XA}& \multirow{2}{*}{XC}&\multirow{2}{*}{XC+XB}& \multirow{2}{*}{XM}& \multirow{2}{*}{arch}	&\multirow{2}{*}{comp}&\multirow{2}{*}{opt}&\multicolumn{2}{c} {XM}\\ 
			&&&&&&& && {MRR10}&{Recall@1}\\
			\midrule
			GraphMoco&with StrandCNN	& \textbf{0.88}&\textbf{0.88}&\textbf{0.89}&\textbf{0.89}& \textbf{0.99}&0.81&\textbf{0.99}&\textbf{0.60}&\textbf{0.55}\\
			
			\midrule
			GraphMoco&with ASM2vecPlus	& 0.87&0.86&0.87&0.87& \textbf{0.99}&0.79&0.89&0.58&0.53\\
			
			\midrule
			GraphMoco&opc 200	& 0.63&0.84&0.81&0.72& 0.67&\textbf{0.86}&0.91&0.24&0.18\\
			\bottomrule[1.5pt]
			\label{tab1}
	\end{tabular}}
\end{table}


\section{Discussion \& Conclusion}
Our objective is to develop a model that can effectively operate in real-world scenarios while maintaining a high level of accuracy. To address the need for both efficacy and accuracy, we have designed a streamlined CNN-based encoder called StrandCNN. This encoder is specifically tailored to handle large volumes of data efficiently.
In conjunction with the StrandCNN encoder, we have devised a graph-based momentum contrastive learning model that leverages the preshuffle mechanism. This approach effectively addresses the issue of shortcuts learned during pretext tasks. By incorporating structural features at various levels, our model achieves SOTA performance on multiple datasets across multiple evaluation metrics. Notably, our model exhibits advantages in terms of both memory usage and computational complexity.

We hope that our research and findings will serve as a source of inspiration for other researchers in related field. By presenting an effective and efficient solution for real-world applications, we aim to contribute to the ongoing advancements in this domain.

\bibliographystyle{cas-model2-names}


\bibliography{mybibfile.bib}


\end{document}